\documentclass{article}
\usepackage{arxiv}

\usepackage[utf8]{inputenc} 
\usepackage[T1]{fontenc}    
\usepackage[colorlinks=true, linkcolor=black, urlcolor=blue, citecolor=blue, anchorcolor=blue]{hyperref}      
\usepackage{url}            
\usepackage{booktabs}       
\usepackage{amsfonts}       
\usepackage{nicefrac}       
\usepackage{microtype}      
\usepackage{lipsum}		
\usepackage{todonotes}
\usepackage{indentfirst}
\usepackage{authblk}

\usepackage{wrapfig}
\usepackage{comment}
\usepackage{textgreek}
\usepackage{natbib}
\usepackage{doi}
\usepackage{enumitem, comment, xifthen}
\usepackage{amsmath,amssymb,amsthm, amsfonts}
\usepackage{makecell}
\usepackage{booktabs}
\usepackage{todonotes}
\usepackage{indentfirst}
\usepackage{setspace}
\usepackage{soul}
\usepackage{float}
\usepackage{hhline}
\usepackage{array,multirow}
\usepackage{lipsum}
\usepackage{siunitx,etoolbox}
\usepackage{graphicx}

\usepackage{tikz}
\usepackage{lipsum}
\usepackage{subcaption} 

\title{Preference-based learning for news headline recommendation}

\author[1]{\textbf{Alexandre Bouras}}
\author[1,2]{\textbf{Audrey Durand}}
\author[1]{\textbf{Richard Khoury}}
\affil[1]{IID – Université Laval}
\affil[2]{Mila – Quebec AI Institute}

\date{}

\renewcommand{\headeright}{}
\renewcommand{\undertitle}{}
\renewcommand{\shorttitle}{Preference-based learning for news headline recommendation}

\hypersetup{
pdftitle={Preference-based learning for news headline recommendatio},
pdfauthor={Alexandre Bouras, Audrey Durand, Richard Khoury},
pdfkeywords={Recommender Systems, Contextual Bandits, Preference Learning},
}

\begin{document}
\maketitle

\begin{abstract}
    This study explores strategies for optimizing news headline recommendations through preference-based learning. Using real-world data of user interactions with French-language online news posts, we learn a headline recommender agent under a contextual bandit setting. This allows us to explore the impact of translation on engagement predictions, as well as the benefits of different interactive strategies on user engagement during data collection. Our results show that explicit exploration may not be required in the presence of noisy contexts, opening the door to simpler but efficient strategies in practice.
\end{abstract}


\section{Introduction}
Social media have become the primary gateway for news consumption. In this setting, user engagement with articles follows a power-law distribution: a small fraction of headlines garners most of the attention. This emphasizes the critical role of headlines in capturing user interest~\citep{nechushtai2019kind}. It thus becomes important to distinguish which headline will attract the most attention, for instance to choose between two different headlines for an article or to decide which headline to share on a social media. However, capturing user engagement is difficult due to behavioural complexity, demographic diversity, and external factors like timing and competition~\citep{o2016theoretical}. 

Moreover, while learning to predict the engagement score might be appealing, this can lead to over-optimization on the evaluation metric~\citep{goodhart1984problems}. This has led to the adoption of preference-based learning~\citep{akrour2011preference} for fine-tuning LLMs~\citep{ouyang2022training}.

In this paper, we leverage pairwise ranking~\citep{koppel2020pairwise} to model user preferences, with the aim of learning a model that can directly rank headlines without predicting user engagement scores (Sec.~\ref{sec:pref_reco}). We further study the performance of our approach under two key challenges in news headline recommendation (Sec.~\ref{sec:expes}). First, we evaluate the impact of language by comparing original French headlines with their English translations in a supervised learning setting. Second, we investigate the impact of learning the model through interactions within a contextual bandit framework~\citep{li2010contextual} rather than using a precollected dataset. Our results (Sec.~\ref{sec:results}) show that preference-based learning performs robustly across languages, while adapting to real-world constraints of model-dependent data collection and delayed feedback.

\newpage
\section{Preference-based Headline Recommendation}
\label{sec:pref_reco}

Preference-based learning proceeds through relative comparisons rather than directly predicting scalar scores. In this case, rather than learning a predictive model for some user engagement measure (e.g., click counts), our goal is to learn a model capable or ordering a set of headlines according their interaction measure. Inspired by approaches used in pairwise ranking and preference models~\citep{durand2018machine, koppel2020pairwise, ouyang2022training}, we propose to train a predictive model $f$ to associate a preference score $f(x)$ to a given input (headline) $x$, such as to preserve the ordering in the user engagement space. Formally, let $x$ and $x'$ denote two inputs with user engagement scores $y$ and $y'$ respectively, if $y > y'$, then we should have $f(x) > f(x')$.

In order to learn this preference model, we consider the Margin Ranking Loss (MRL)~\citep{sculley2009large}:

\begin{equation}
\label{eq:MRL}
    \mathcal{L}(x, x') := \max(0, m - (f(x) - f(x')) \cdot p(x, x')),
\end{equation}

where $p(x, x') := \operatorname{sgn}(y - y')$ denotes the preference direction and $m \geq 0$ is a tolerance margin. The margin $m$ balances sensitivity and generalization, with smaller values capturing finer distinctions and larger values improving robustness.

\subsection{Pairwise Data Preparation}
\label{sec:pairwise_prep}

Given $n$ data points $\{ (x_i, y_i) \}_{i = 1\dots n}$, we use pairwise rankings to generate training and evaluation datasets. First, the headlines are grouped into engagement ranks using a binning function that discretizes the user engagement space. Note that these bins do not need to be of equal size; previous research has shown that logarithmic binning is especially useful when discretizing power law distributions~\citep{milojevic2010power} such as user engagement. Let $z_i$ correspond to the rank of headline $x_i$. Each headline is then paired with $M$ headlines sampled from all superior ranks, creating pairs $(x_i, x_j)$, s.t. $z_i < z_j$ (thus $y_i < y_j$). For example, consider headlines $x_1$, $x_2$, and $x_3$ with respective ranks $z_1 = 0$, $z_2 = 1$, and  $z_3 = 2$. With $M = 1$, these would result into the dataset $\mathcal D = \{(x_1, x_2), (x_1, x_3), (x_2, x_3)\}$. Pairing based on bins rather than engagement scores ensures that each article is compared to a set of articles representing the complete range of engagement values, and prevents noisy comparisons in the form of same-bin comparisons.

\section{Experiments}
\label{sec:expes}

All experiments are repeated with 100 random seeds to introduce stochasticity in model initialization, pairwise data preparation, and interactive learning behaviour.

\subsection{Data}

The data consists of Facebook posts published by \textit{Le Soleil}, a regional French Canadian newspaper. Each post contains a news headline and is associated with user engagement metrics, with click count used as the primary metric. The dataset spans from November 1, 2019, to September 23, 2021, and includes a total of 3,305 posts~\citep{bouras2024impact}. 
Click counts are cumulated over a 7-day period after posting to account for observed user interaction trends, where most clicks occur within the first week of publication. 
This data is divided into training (80\%) and test (20\%) data, using chronological order to respect the real-world setting.

\paragraph{\textbf{Data preparation}}

Table~\ref{tab:binning_rule} shows the binning function we used to determine engagement ranks. This function was developed in collaboration with \textit{Le Soleil}, ensuring the intervals reflect meaningful engagement distinctions relevant to the publication’s content strategy.
\begin{table}[H]
\centering
\begin{tabular}{|l|c|c|c|c|c|c|c|}
    \hline
    Rank & 0 & 1 & 2 & 3 & 4 & 5 & 6\\
    \hline
    \hline
    Min. bound (incl.) &  0 & 100 & 1000 & 5000 & 10,000 & 50,000 & 100,000\\
    \hline
    Max. bound (excl.) & 100 & 1000 & 5000 & 10,000 & 50,000 & 100,000 & $\infty$\\
    \hline
    Number of headlines &883 &1660 &583 &96 &60 &19 &4 \\
    \hline
\end{tabular}
\caption{Bounds defining each rank, with the number of headlines per rank.}
\label{tab:binning_rule}
\end{table}

\subsection{Evaluation Metrics}

Let $\mathcal D$ denote an evaluation dataset of headline pairs and let $\hat p(x, x') := \operatorname{sgn}\big(f(x) - f(x')\big)$ denote the predicted preference. We evaluate the performance of a preference model using two key metrics, where $\mathbb I [\text{condition}]$ is the indicator function.
The \textit{accuracy} measures the proportion of correctly-ordered pairs:
\begin{equation}
\label{eq:accuracy}
    \text{accuracy} := \frac{\sum_{(x, x') \in \mathcal D} \mathbb I[\hat p(x, x') = p(x, x')]}{|\mathcal D|}.
\end{equation}
It reflects the model’s ability to capture relative preferences across all ranks.
The \textit{weighted accuracy} accounts for the fact that ranks are not equally represented in $\mathcal D$. Let $K$ denote the number of ranks and $\mathcal D^{(k)} \subseteq \mathcal D$ denote the subset of pairs containing at least one item of rank $k$. The weighted accuracy averages the proportion of correctly ordered pairs \textit{per rank}:
\begin{equation}
\label{eq:weighted_accuracy}
    \text{weighted accuracy} := \frac{1}{K} \sum_{k=0}^{K-1} \frac{\sum_{(x, x') \in \mathcal D^{(k)}} \mathbb I[\hat p(x, x') = p(x, x')]}{|\mathcal D^{(k)}|}.
\end{equation}
It ensures a balanced evaluation of ordering performance across all levels of user engagement.

\subsection{Learning in Different Languages}

Language can introduce variability in performance, as lower-resource-language embeddings such as French may differ in their robustness and representation quality compared to the ones of data-rich languages such as English.
We therefore compare our results with the original French headlines with their English translations\footnote{Generated by DeepL: \url{http://wwww.deepl.com},, which has demonstrated high accuracy and reliability for translation quality~\citep{yulianto2021google}.}. Each headline is represented as a feature vector derived from sentence embeddings generated by NV-Embed~\citep{lee2024nv} and BGE-Multilingual-Gemma2~\citep{li2024making}, which respectively process English and French headlines. 

\paragraph{\textbf{Evaluation setting}} We consider the typical supervised learning scenario, assuming that the training data has been pre-collected and test data simulates what would be seen at deployment. The training data is further split in data actually used for training (90\%) and data used for validation (10\%) during the learning process. The validation set is used for early stopping and learning rate adjustments during training to prevent overfitting. Pairwise data preparation (Sec.~\ref{sec:pairwise_prep}) is conducted with $M = 2$ samples.

\subsection{Online Learning through Interactions}

To study the impact of interactive learning, we use all data to simulate an online recommendation system under the contextual bandit framework with delayed feedback~\citep{chapelle2011empirical}. We assume a 90-days warm-up phase to avoid cold-start issues~\citep{mary2014bandits}. This is simulated by considering an initial history $\mathcal H_0$ of 90 headlines randomly sampled from the the first 90 days of the training data. History $\mathcal H_0$ produces dataset $\mathcal D_0$ using pairwise data preparation (Sec.~\ref{sec:pairwise_prep}) with $M=2$, and is used for training an initial preference model $f_0$. The remaining headlines in the data are grouped into sets of available headlines $\mathcal X_t$ for each time (day) $t=1,2,\dots,T$, with $T=485$. Then, for each time $t$, the recommender agent selects headline $X_t \in \mathcal X_t$ using preference model $f_{t-1}$. The associated click count $Y_t$ is observed at a delayed time $t+7$, to simulate the agent waiting 7 days before observing the impact of their actions, aligning with real-world engagement patterns~\citep{chapelle2014modeling}. When the agent receives the feedback $Y_{t'}$ associated with a previous recommendation $X_{t'}$, its history is updated, i.e. $\mathcal H_t \leftarrow \mathcal H_{t-1} \cup \{ (X_{t'}, Y_{t'}) \}$, and dataset $\mathcal D_t$ is generated for training model $f_t$. If no feedback is received on time $t$ (e.g. for all $t < 7$), then $\mathcal H_t \leftarrow \mathcal H_{t-1}$, $\mathcal D_t \leftarrow \mathcal D_{t-1}$, and $f_t = f_{t-1}$.

\paragraph{\textbf{Online performance}} Let $Y_t$ denote the number of clicks received for the headline recommended at time $t$, $Y_{t}^\star$ denote the number of clicks for the \textit{best} headline available at time $t$, and $Y_{t}^-$ denote the number of clicks for the \textit{worst} headline available at time $t$. The \textit{total clicks} measures the number of clicks received with the recommended headlines over $T$ days:
\begin{align}
    \label{eq:total_clicks}
    \text{total clicks} := \sum_{t=1}^T Y_t.
\end{align}
The \textit{normalized total clicks} evaluates the outcome relative to the maximum and minimum achievable outcomes, to account for imbalances in click counts across publications:
\begin{equation}\label{eq:norm_clicks}
    \text{normalized clicks} := \sum_{t=1}^{T} \frac{(Y_t - Y_{t}^-)}{(Y_{t}^\star - Y_{t}^-)}.
\end{equation}

\paragraph{\textbf{Preference model evaluation}} Accuracy (Eq.~\ref{eq:accuracy}) is assessed for each newly-trained model $f_t$, as long as $t \leq 335$, when the online simulation begins including test data.

\paragraph{\textbf{Strategies}} We consider consider Neural Thompson Sampling (NeuralTS)~\citep{zhang2021neural} and a greedy selection. NeuralTS uses posterior sampling to balance exploration and exploitation in online learning, while the greedy strategy selects $X_t = \arg\max_{x \in \mathcal X_t} f_{t-1}(x)$ (i.e. no exploration).

\paragraph{\textbf{Baselines}} To evaluate the impact of agent choices on collected data, we compare to a random headline selection. As reference point, we also consider the best choice and second-best choice oracles (unavailable in practice).

\subsection{Model Architecture}

We consider a neural network with two fully-connected layers and ReLU activations, residual connections, batch normalization, and a final ReLU activation. Residual blocks were chosen for their ability to act as ensembles of networks, improving both robustness and generalization. 
This architecture and hyperparameters were optimized using Optuna~\cite{akiba2019optuna}\footnote{With the Ray Tune library: \url{https://docs.ray.io/}.}. 
The search space included learning rates in \{0.001, 0.005, 0.01, 0.05, 0.1\}, weight decay in $\{0.0001, 0.001, 0.01\}$, number of residual blocks in \{1, 2, 3\}, neurons per layer in \{100, 200, 300, 400\}, MRL (Eq.~\ref{eq:MRL}) margin $m \in \{0.0, 0.5, 1.0\}$, and learning rate patience \{0, 1, 2\}. The Adam optimizer was configured with $\epsilon = 10^{-8}$,  $\beta_1 = 0.9$, and $\beta_2 = 0.999$. 
Optimization was conducted using French and English embeddings of dimensionality of 1536 (NV-Embed) and 1024 (BGE-Multilingual-Gemma2), respectively.

\section{Results}
\label{sec:results}

The optimized model consists of a single residual block with 200 neurons. The learning rate, initialized at 0.005, is dynamically reduced by a scheduler after one epoch without improvement in validation loss. The model is trained using a batch size of 128, a weight decay of 0.001, and the MRL (Eq.~\ref{eq:MRL}) with a margin $m=1$.

\subsection{Language of Headline Embeddings}
\label{sub:embeddings}
For French embeddings, the model achieved an average accuracy of $84.48$\% ($\pm1.90$) and a weighted accuracy of $84.86$\% ($\pm2.39$). English embeddings showed a similar performance, with an average accuracy of $85.13$\% ($\pm3.37$) and a weighted accuracy of $85.19$\% ($\pm4.42$). Although both embeddings show similar accuracy and weighted accuracy results, the English-based model exhibits slightly greater variance. This is likely due to the translation, which can introduce stochastic effects in the embedding process. This shows that using original-language embeddings is preferable. But, if these embeddings are not available, as is the case for many lower-resource languages, then translating to English and using embeddings in that language can be done with little degradation in the results.

\subsection{Impact of Learning through Interactions}
\label{sub:inter_learning}

\begin{figure}
    \centering
    \begin{subfigure}{.48\textwidth}
    \centering
         \includegraphics[width=\textwidth, height=4.7cm]{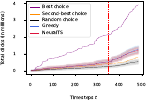}
         \caption{Cumulative clicks.}
         \label{fig:clicks}
    \end{subfigure}
    \hfill
    \begin{subfigure}{.48\textwidth}
    \centering
         \includegraphics[width=\textwidth, height=4.7cm]{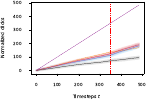}
         \caption{Cumulative normalized clicks.}
         \label{fig:normalized_clicks}
    \end{subfigure}
    \caption
    {Average online performance of interactive strategies and baselines.
    }
    \label{fig:inter_learning}
\end{figure}

Figure~\ref{fig:inter_learning} shows the online performance of interactive strategies and baselines. We observe that the NeuralTS and greedy strategies both significantly outperform random selection in terms of total clicks (Eq.~\ref{eq:total_clicks}) cumulated over time and normalized clicks (Eq.~\ref{eq:norm_clicks}), highlighting their benefits on user engagement. We also observe an important difference between the results of the best and second-best choice oracles in both metrics, highlighting the power-law dynamics in click distributions (i.e. the drop in engagement between the first and second-most popular headlines). We also observe that both NeuralTS and greedy strategies eventually (around time $t=350$) overcome the second-best choice oracle. Interestingly, the expected advantage of exploration in NeuralTS compared to the greedy strategy is not observed. This can be attributed to the limited granularity of preference-based context representations causing implicit exploration, along with the exploration contained in the 90-day warm-up phase~\citep{kannan2018smoothed, bayati2020unreasonable}. Consequently, the explicit exploration of NeuralTS may not be required, and simpler strategies such as greedy selection can be as efficient in practice.

\begin{figure}
    \centering
     \includegraphics[width=0.58\textwidth]{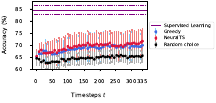}
    \caption
    {Evolution of model accuracy over time.}
    \label{fig:superv_learning_comp}
\end{figure}

Figure~\ref{fig:superv_learning_comp} shows the accuracy over time of the preference model on the test set  during online learning. We include the model trained on all training data ($2,644$ headlines), while the online models are trained on 425 headlines. We observe that the online models already have a large variance at $t=0$ (due to both model initialization and warm-up data-sampling), and it further increases over time. This highlights a key challenge of interactive learning: each recommendation is heavily influenced by previous ones, and suboptimal recommendations early on can lead to the selection of data that does not improve the model, compounding the effects on the overall performance. However, interactive data collection (e.g. NeuralTS and greedy) lead to better preference models on average than random data collection (i.e. supervised learning with same number of headlines as interactive strategies), since it allows information acquisition on more relevant headlines.

\section{Conclusion}

This work addresses the challenge of recommending news headlines on social media, where user engagement follows a power-law distribution. We propose a preference-based learning framework using pairwise ranking to model relative preferences, to avoid the over-optimization risks that stem from using engagement metrics.
Our results show similar performances in French and English, demonstrating the potential of translation to leverage available models.

\newpage
\bibliographystyle{unsrtnat}
\bibliography{references}

\end{document}